\documentclass[12pt,a4paper]{article}
\usepackage{graphicx}
\usepackage[dvips]{epsfig}
\newcounter{fig}

\title{Electron-phonon interaction and coupled phonon--plasmon modes}

\author{L. A. Falkovsky\\
Landau Institute for Theoretical Physics, 117337 Moscow, Russia}

\bigskip
\bigskip
\begin{document}
\maketitle

\begin{abstract}
The theory of Raman scattering by the electron--phonon coupled system
in metals and heavily doped semiconductors  is developed 
taking into account the Coulomb screening and the   electron--phonon deformation
interaction. The Boltzmann equation for carriers is applied. Phonon
frequencies and optic coupling constants are renormalized due to interactions
with carriers. The $k-$dependent semiclassical  dielectric function is involved
instead of the Lindhard-Mermin expression. The results of  calculations
are presented for various values of carrier concentration and electron-phonon
coupling constant. 

PACS number(s): 63.20.Dj, 63.20.Kr, 71.38-k, 72.30.+q, 78.30.-j 
\end{abstract}

\vskip .5cm
\section{Introduction}

Recently  there has been considerable interest in the effect of
electron--phonon interactions on the optical-phonon dispersion. 
This interest is stimulated by  contradictions
between different approaches to the electron-phonon interaction. 
The strong phonon renormalizations were obtained first by Migdal \cite{Mi}
(see also \cite{AGD}) within a consistent many-body approach based on 
the Fr\"olich Hamiltonian.
The extremely large dispersion of optical phonons was  predicted in 
work \cite{AS} using also the Fr\"olich model. 
These results contradict to the Born--Oppenheimer (adiabatic)
concept \cite{BHK} according to which the phonon renormalizations 
should be small
in terms of the nonadiabatic parameter $\sqrt{m/M}$, where $m$
and $M$ are the electron and ion masses, respectively  (see, also \cite{BK}).
 Theoretical
investigations \cite{Kon} of the sound velocity  and
 acoustic attenuation in metals confirm the adiabatic concept. 
In a recent paper, Reizer  \cite{R} emphasized the importance 
of screening  which
should be taken into account. To our knowledge, 
 the Coulomb screening effect on the LO phonons was firstly studied 
 in the work \cite{GLF}. In our work \cite{Fa}, using the Boltzmann equation
 we have found that the electron-phonon interaction results more essentially
 in the optical--phonon damping than in the dispersion law.  In any case, 
  the Fr\"olich model has the evident shortcomings.
 
 From the experimental point of view, the best opportunity for the 
 investigation
 of  interactions between  electrons and optical phonons
is provided by coupled phonon-plasmon modes in doped semiconductors
(see, e.g. \cite{ACP}). 
Two such  modes $L^{\pm}$ have been  observed
in Raman experiments for many semiconductors. 
On the early step,   the Raman results
 were compared with the
theory \cite{HF} based  on a Drude model (see,  e.g. \cite{BIM}), but
more recently the Lindhard-Mermin expression for the dielectric function
 has been used \cite{ACI}.

The Lindhard-Mermin expression \cite{LM} 
represents a sophisticated generalization 
of the Lindhard function  with the help of the electron relaxation time.
The Lindhard approach is very useful while the momentum transfer $k$ in
the Raman scattering  is compared with the Fermi momentum $p_F$.
The most essential influence of
carriers should be expected for $kv_F\sim \omega$, where $v_F$ is the Fermi
velocity and $\omega$ is the phonon frequency. For solids with metalic
conductivity, the Fermi velocity could be estimated using an argument of
stability under the Coulomb interaction $ e^2/\pi \hbar v_F\le 1$. This
condition gives $v_F\sim 0.7\times 10^8$ cm/s. For the typical value
of the optical phonon frequency $\omega = 500$ cm$^{-1}$, the interesting
values $k\le\omega/v_F\simeq 10^6$ cm$^{-1}$. 
Therefore, the condition $k<p_F$ is
satisfied for the carrier concentration larger than 3$\times 10^{17}$cm$^{-3}$.
In experiments,  
the heavily doped semiconductors with  large carrier concentration are used,
in order to obtain a visible effect of carriers. 
In such a situation, the condition $ k\ll p_F$ is fulfilled, 
and we can  apply the Boltzmann equation in  calculations of
electronic susceptibility, as well as  for the evaluation of
 the Raman cross section.
The method of the Boltzmann equation is valid for 
  the anisotropic electron plasma in solids at  arbitrary temperatures. 
  In the present paper, 
we  obtain the Raman efficiency applying the Boltzmann equation for 
 generate carriers in heavily doped semiconductors when  temperature 
is lower than the Fermi energy, $T\ll \varepsilon_F$.

\section{Effective Hamiltonian and light scattering}  

For the electron--phonon system in solids, we use  the operator of 
particle numbers $\hat{n}$, the phonon displacements $\hat{b}_j$, 
and the macroscopic
electric field $E$, accompanying vibrations in polar semiconductors 
and acting on charges of electrons and ions.
The effective Hamiltonian describing the inelastic light scattering in
solids can be written in the semiclassical Wigner representation
 \begin{equation} \label{hami}
{\cal H} =
 {e^{2} \over m c^{2}}
 \int d^{3}r\,
 {\cal N}({\bf r},t)U({\bf r},t),
 \end{equation}
where
\begin{equation} \label{1a}
{\cal N}({\bf r},t) =
\gamma\hat{ n}({\bf r},t)+
g_{j} {\hat b}_j({\bf r},t)+
g_E E({\bf r},t)
\end{equation}
is a linear form of variables $\hat{n}$,  $\hat{b}_j$, and $E$.
The subscript $j$ denotes the various phonon modes: longitudinal (LO)
or transverse (TO). More precisely, the subscript $j$ numerates
the different phonon representations which can be degenerate. The
 transformation properties of the coupling constants $g_{j}$ 
 are determined by this
representation.
The notation $U({\bf r}, t)$ is introduced for a product of the
vector-potentials of incident  and scattered  photons:
$$ A^{(i)}({\bf r},t) A^{(s)}({\bf r},t)=
U({\bf r}, t)=
\exp[i({\bf k}\cdot{\bf r}- \omega t)]
U({\bf k}, \omega),$$
where the momentum and frequency transfers 
${\bf k} = {\bf k}^{(i)} - {\bf k}^{(s)}$,
$\omega = \omega^{(i)} - \omega^{(s)}$. The polarization vectors of 
${\bf \hat b}_j({\bf r}, t)$,  ${\bf E}({\bf r}, t)$,  
$ {\bf A}^{(i)}({\bf r},t)$, and 
$ {\bf A}^{(s)}({\bf r},t)$ are included in
the coupling constants.

 The first term on the right-hand side of Eq. (\ref{1a})
 descibes the light scattering by electron--hole pairs 
with a vertex
$$\gamma({\bf p}) = e^{(i)}_{\alpha }e^{(s)}_{\beta
}\left[\delta_{\alpha \beta } + {1\over m}\sum_{n}^{}\left({p^{\beta
}_{fn} p^{\alpha }_{nf}\over\epsilon _{f}({\bf p}) - \epsilon _{n}({\bf
p}) + \omega^ {(i)}} + {p^{\beta }_{fn}p ^{\alpha }_{nf}\over \epsilon
_{f}({\bf p}) - \epsilon _{n}({\bf p}) - \omega^{(s)}}\right)\right]$$
where the resonant term is included; 
$e^{(i)}_{\alpha }$ and $e^{(s)}_{\beta}$ are
the polarizatin vectors of incident $ {\bf A}^{(i)}({\bf r},t)$ and scattered 
$ {\bf A}^{(s)}({\bf r},t)$ photons. 
The quantum-mechanical and statistical average of the first term in Eq. 
(\ref{1a})
\begin{equation} \label{ng}
\langle\langle\gamma \hat{n}({\bf r},t)\rangle\rangle
 = \int {2d^{3}p\over (2\pi ) { }^{3}}
\gamma ({\bf p})f_{p}({\bf r},t) 
\end{equation}
can be expressed in terms of the electron distribution function 
$f_{p}({\bf r},t)$.
 The   constants $g_{j}$ and  $g_E$ 
are the deformation-optic
 and electro-optic  couplings with the phonon 
displacements and the macroscopic electric field, correspondingly. The
estimation gives $g_j\sim /a^4$, 
$g_{E}\sim 1/ea$, and $\gamma({\bf p})\sim m^{*}/m$,
where $a$ is the lattice parameter and $m^{*}$ is the effective mass.

The variable  $U({\bf r}, t)$ can be considered as an external force.
Then one  introduces the generalized  susceptibility $\chi ({\bf k}, \omega )$
 as the linear response to this force:
 \begin{equation} \label{11}
\langle\langle{\cal N}({\bf k},\omega)\rangle\rangle
 \,= - \chi ({\bf k}, \omega )
U({\bf k} , \omega).
\end{equation}
According to the fluctuation-dissipation theorem  the function
$$K({\bf k}, \omega )
= {2\over 1 - \exp(-\omega /T)}~ {\rm Im}\, 
\chi ({\bf k}, \omega )$$
is the Fourier component of the correlation function
 \begin{equation} \label{9}
K({\bf r},t;{\bf r}^{\prime},t')
=\langle\langle {\cal N}^{\dag}({\bf r},t)
{\cal N}({\bf r}^{\prime},t')\rangle\rangle
\end{equation}
which depends only on the differences ${\bf r-r^{\prime}}$ and
$t-t^{\prime}$.
The  Raman cross section  is given by 
\begin{equation}
\frac{d\sigma}{d\omega^{(s)} d\Omega^{(s)}}=
{\frac{ k_{z}^{(s)} \omega^{(s)}}{\pi c}} \left( \frac{2e^{2}}
{ c \hbar m \omega^{(i)}}\right)^{2}  
K({\bf k},\omega)
|U({\bf k},\omega)|^2,
\label{8}
\end{equation}
where $k_{z}^{(s)}$ is the normal to the sample surface component
of the scattered wave vector in vacuum.

One note should be made. Of course, any sample has  the surface. The surface 
effects in the Raman 
scattering was considered in our work \cite{FM} and they are  omitted in the
derivation of Eg. (\ref{8}). Furthermore, the incident and scattered fields
do not penetrate into the bulk due to the skin effect. For the optical range
of the incident light, we have the normal skin-effect conditions. Then we
 integrate in Eq. (\ref{8}) the distribution $|U({\bf k},\omega)|^2$ 
over the normal component $k_z$. As shown in the paper \cite{FM},
 the integration of  $|U({\bf k},\omega)|^2$  gives a factor $1/\zeta_2$, 
 where $\zeta_2$ is expressed in terms of
the wave-vector components {\it inside semiconductor}:
$\zeta_2= {\rm Im} (k_z^{(i)}+
k_z^{(s)})$. The obtained Raman cross section (\ref{8}) is dimensionless. 
It represents a ratio
of the inelastic scattered light energy  to the incident energy.

\section{Boltzmann equation for carriers}

The problem of the evaluation of the Raman cross section  consists in the
calculation of the generalized susceptibility (\ref{11}). 
We apply the Boltzmann equation for the electron distribution function:
\begin{equation}\label{be}
\frac{\partial f_{p}({\bf r},t)}{\partial t}+ {\bf v}
\frac{\partial f_{p}({\bf r},t)}{\partial {\bf r}}+
\dot{\bf p}\frac{\partial f_{p}({\bf r},t)}{\partial {\bf p}}=
-\frac{1}{\tau}[f_{p}({\bf r},t)-\langle f_{p}({\bf r},t)\rangle].
\end{equation}
The angular brackets denote the average over the Fermi surface
$$\langle...\rangle = \frac{1}{\nu_0}\int(...){2dS_F\over v(2\pi)^3},$$
where  the integral is performed in the momentum space
over the Fermi surface  
and $\nu_0$ is the density of electron
states, defined by the condition $\langle 1 \rangle=1$.  
We use the $\tau-$approximation which is correct for
the electron scattering by defects in metals and heavily 
doped semiconductors. The collision integral in the form (\ref{be}) 
conserves the number of electrons
in collisions. Therefore, the charge density satisfies  the   
equation of continuity. This ensures the correct $\omega-$dependance 
of the dielectric function at low frequencies.

According to Eqs. (\ref{hami}),  (\ref{1a}) and (\ref{ng}) 
instead of the unperturbed electron spectrum  $\varepsilon_0({\bf p})$,
we introduce the local electron  spectrum in 
the presence of the external force
$U({\bf r}, t)$ as follows:
$$ \varepsilon({\bf p}, {\bf r},t)=\varepsilon_0({\bf p})+ 
\gamma ({\bf p})U({\bf r}, t)
 +\zeta_j({\bf p}) b_j({\bf r},t),$$
where the last term represents the electron--optical-phonon deformation
potential and $b_j({\bf r},t)=\langle\langle\hat b_j({\bf r},t)\rangle\rangle$.
We use this form of the el-ph interaction instead of the  Fr\"{o}lich  
poirization type $\zeta({\bf p}) {\rm div} {\bf b}({\bf r},t)$, 
because it is larger by the parameter $1/ka$ for the optical phonons.

Let us linearize Eq. (\ref{be}), looking for its solution in the form
\begin{equation}\label{lbe}
f_p({\bf r},t)=f_0[\varepsilon({\bf p, r},t)-\mu] - \frac{df_0}{d\varepsilon}
\delta f_p({\bf r},t),
\end{equation}
where $f_0[\varepsilon({\bf p, r},t)-\mu]$ is the Fermi--Dirac local
distribution function.
It is important that the collision term in the Boltzmann equation is canceled by
the local-equilibrium term of Eq. (\ref{lbe}). 

We set the number-conservation condition on the chemical potential:
$$ \int\frac{d^3p}{(2\pi)^3}f_0[\varepsilon({\bf p,r},t)-\mu]=
\int\frac{d^3p}{(2\pi)^3}f_0(\varepsilon_0-\mu_0).$$
We obtain
$$\mu=\mu_0+\langle\gamma({\bf p})\rangle U({\bf r}, t)
 +\langle\zeta_{j}({\bf p}) \rangle b_j({\bf r},t).$$
 The condition implies the renormalization of vertices
 \begin{equation}\label{scr}
 \gamma({\bf p})\rightarrow\gamma({\bf p})-\langle\gamma({\bf p})\rangle,\quad
 \zeta_{j}({\bf p})\rightarrow\zeta_{j}({\bf p})-\langle\zeta_{j}({\bf p})\rangle
\end{equation}
and this substitution should be made in  that follows.

The linearized Boltzmann equation  in the Fourier components has the form
$$-i(\omega-{\bf k\cdot v}+i/\tau)\delta f_{p}({\bf k},\omega)=
\psi_p({\bf k}, \omega)+
\langle\delta f_{p}({\bf k},\omega)\rangle/\tau,$$
where
$$\psi_p({\bf k}, \omega)=e{\bf v\cdot E}({\bf k}, \omega)-
i\omega[\gamma({\bf p}) U({\bf k},\omega)+\zeta_{j}({\bf p})b_j({\bf k}, \omega)].$$
The solution to the equation is esily obtained:
\begin{equation} \label{sol}
\delta f_{p}({\bf k},\omega)=i[\psi_p({\bf k}, \omega)
+\langle\delta f_{p}({\bf k},\omega)\rangle/\tau]/\Delta_p,
\end{equation}
where we denote $\Delta_p=\omega-{\bf k\cdot v}+i/\tau$.
Now we get
\begin{equation} \label{sbe2}
\langle \delta f_{p}({\bf k}, \omega)\rangle
=i\langle \psi_p({\bf k},\omega)/\Delta_p\rangle
/(1-i\langle\tau^{-1}/\Delta_p\rangle).
\end{equation}
Notice, that in  accordance with the adiabatic concept,
no additional contribution comes from  the local equilibrium  distribution
function $f_0[\varepsilon({\bf p, r},t)]$ in Eq. (\ref{lbe}). 

\section{Equation of motion for phonons interacting with carriers}

In the
long-wave approximation ($k\ll 1/a$, $a$ being the lattice parameter),
 we write the equation of motion for the phonon displacement field
\begin{equation}
\label{oeq}
(\omega_k^2-\omega ^{2})b_j({\bf k},\omega)
=\frac{Z}{M'}E_j({\bf k},\omega)-\frac{g_{j}U({\bf k}, \omega)}{M'N}
-\frac{1}{M'N}\int{2dS_F\over v(2\pi)^3}
\zeta_ {j}({\bf p})\delta f_{p}({\bf k},\omega),
\end{equation}
where $N$ is the number of unit cells in 1 cm$^3$, 
 $M'$ is the reduced mass of the unit cell, and $Z$ is the effective ionic
 charge. The nonperturbed phonon frequency $\omega_k$ should be considered 
 in the
 absence of the electric field and without any electron--phonon interactions. 
In the long-wave limit, we can expand it as $\omega_k^2=\omega^2_0\pm s^2k^2$
with the value of the dispersion parameter $s$ 
 on the order of the typical sound velocity
in solids. Notice that the optical phonons always have the so-called natural
width $\Gamma^{\rm nat}\sim \omega_0\sqrt{m/M}$. The natural width results from
decay processes into two or more acoustic and optical phonons. In the final
expressions, we will substitute 
$\omega_k^2-\omega^2\rightarrow\omega_k^2-i\omega\Gamma^{\rm nat}-\omega^2$.

The equation (\ref{oeq}) is applied to both the longitudinal and transverse
 phonons.
It is seen from the Maxwell equations, that the electric field is longitudinal
${\bf E}\parallel{\bf k} $
 in the optical region $k\gg\omega/c$. If the excited phonons
propagate in the symmetrical direction, the TO and LO phonons are separated. 
Therefore, the electric field stands only in Eq. (\ref{oeq})  
for the LO phonon. Beside of that, the coupling $\zeta_{j}({\bf p})$  depends on
the phonon representation $j$.

Using the solution (\ref{sol}), we rewrite Eq. (\ref{oeq}): 
\begin{equation}\label{oeq1}
(\tilde{\omega}_{j}^2-\omega ^{2})b_j({\bf k},\omega)
-\frac{\tilde{Z}}{M'}E_j({\bf k},\omega)
=-\frac{\tilde{g}_{j}U({\bf k},\omega)}{M'N},
\end{equation}
where the phonon frequency 
\begin{equation}\label{shd}
\tilde{\omega}_{j}^2=\omega_k^2
+\frac{\omega\nu_0}{M'N}
\left(\big\langle \frac{\zeta_{j}^2({\bf p})}{\Delta_p}\big\rangle
+\frac{i\langle \zeta_{j}({\bf p})/\Delta_p\rangle^2}
{\tau-\langle i/\Delta_p\rangle}\right),
\end{equation}
 the  effective ionic charge 
\begin{equation} \label{oeq2d}
\tilde{Z}=Z-\frac{ie\nu_0}{N}
\left(\big\langle\frac{v_z\zeta_{j}({\bf p})}{\Delta_p}\big\rangle
+\frac{i\langle v_z/\Delta_p\rangle
\langle\zeta_{j}({\bf p})/\Delta_p\rangle}
{\tau-\langle i/\Delta_p\rangle}\right),
\end{equation}
and the deformation-optic coupling
\begin{equation}\label{doc}
\tilde{g}_{j}=g_{j}
+\omega\nu_0
\left(\big\langle \frac{\zeta_{j}({\bf p})
\gamma({\bf p})}{\Delta_p}\big\rangle
+\frac{i\langle \zeta_{j}({\bf p})/\Delta_p\rangle
\langle \gamma({\bf p})/\Delta_p\rangle}
{\tau-\langle i/\Delta_p\rangle}\right)
\end{equation}
are  renormalized due to the electron--phonon 
interaction $\zeta_{j}({\bf p})$.

\section{Poisson equation for the macroscopic field}
 
 Let us consider the longitudinal electric induction $D$ that accompanies the
 lattice vibrations. 
There are several contributions in the  field : 
(1) the polarization $\alpha E({\bf r},t)$ 
of the filled 
electron bands, (2) the lattice polarization   $NZb_{\rm LO}({\bf r},t)$, 
(3) the contribution of free carrier density 
$\rho=-{\rm div}{\bf P}_e$, and (4) the term
$P= -\partial{\cal H}/\partial E= - g_EU$  explicitly results 
from the Hamiltonian (\ref{hami}), (\ref{1a}). 
Collecting all these terms into the Poisson
equation, ${\rm div} {\bf D}=0$, we find
\begin{equation}\label{pe}
\varepsilon_{\infty}E({\bf k},\omega) +4\pi NZb_{\rm LO}({\bf k},\omega)
+\frac{4\pi ie}{k}
\int\frac{2d^3p}{(2\pi)^3}
\delta f_p({\bf k},\omega) -4\pi g_EU({\bf k},\omega)=0,
\end{equation}
where the high-frequency permittivity $\varepsilon_{\infty}=1+4\pi\alpha$.
Using the solution of the Boltzmann equation we rewrite the Poisson equation in
the form
\begin{equation}\label{pe1}
\varepsilon_{e}({\bf k},\omega)E({\bf k},\omega) +4\pi N \bar{Z}
b_{\rm LO}({\bf k},\omega) = 4\pi \tilde{g}_EU({\bf k},\omega),
\end{equation}
where the electronic dielectric function
\begin{equation}\label{dfu1}
\varepsilon_e({\bf k},\omega)=\varepsilon_{\infty}+
\varepsilon_{\infty}\frac{k_0^2}{k^2 }
\left[1-\frac{\langle \omega/\Delta_p(k)\rangle}
{1-\langle i/\Delta_p(k)\rangle/\tau}\right]
\end{equation}
with the Thomas-Fermi parameter
$k_0^2=4\pi e^2\nu_0/\varepsilon_{\infty}$.

Due to the electron-phonon interactions $\zeta_{\rm LO}({\bf p})$, 
the ionic charge  obtains  an additional term
\begin{equation} \label{oeq3d}
\bar{Z}=Z+\frac{ie\nu_0}{N}
\left(\big\langle\frac{ v_z\zeta_{\rm LO}({\bf p})}{\Delta_p}\big\rangle
+\frac{i\langle v_z/\Delta_p\rangle
\langle\zeta_{\rm LO}({\bf p})/\Delta_p\rangle}
{\tau-\langle i/\Delta_p\rangle}\right),
\end{equation}
of the opposite sign in the comparison with that in Eq. (\ref{oeq2d}).
The electro-optic coupling in Eq. (\ref{pe}) also changes, but because of
the light scattering  by carriers $\gamma({\bf p})$:
\begin{equation} \label{oege}
\tilde{g}_E=g_E-ie\nu_0
\left(\big\langle\frac{v_z\gamma ({\bf p})}{\Delta_p}\big\rangle
+\frac{i\langle v_z/\Delta_p\rangle
\langle\gamma({\bf p})/\Delta_p\rangle}
{\tau-\langle i/\Delta_p\rangle}\right).
\end{equation}

\section{ Raman scattering by electron--hole
pairs, phonons, and coupled modes}
 
 Now we are in a position to calculate the susceptibility
 (\ref{11}). Using   Eqs. 
(\ref{ng}), (\ref{sol}), (\ref{sbe2}), (\ref{doc}), and (\ref{oege}), 
we get
\begin{equation}\label{secs}
\langle\langle{\cal N}({\bf k},\omega)\rangle\rangle= 
- \chi_e ({\bf k}, \omega )U({\bf k} , \omega)
+\tilde{g}_jb_j({\bf k},\omega)+\bar{g}_EE({\bf k},\omega),
\end{equation}
where 
$$\chi_e ({\bf k}, \omega )=-\omega\nu_0
\left(\big\langle \frac{\gamma^2({\bf p})}{\Delta_p}\big\rangle
+\frac{i\langle \gamma({\bf p})/\Delta_p\rangle^2}
{\tau-\langle i/\Delta_p\rangle}\right)$$
gives the light scattering with excitation of  electron--hole pairs.
Notice, that here the renormalized coupling $\bar{g}_E$ differs 
from $\tilde{g}_E$  (\ref{oege}) in the sign of
the second term:
$$\bar{g}_E=g_E+ie\nu_0
\left(\big\langle\frac{v_z\gamma ({\bf p})}{\Delta_p}\big\rangle
+\frac{i\langle v_z/\Delta_p\rangle
\langle\gamma({\bf p})/\Delta_p\rangle}
{\tau-\langle i/\Delta_p\rangle}\right).$$

In order to find $E({\bf k},\omega)$ and $b_j({\bf k},\omega)$, 
we have to solve
the system of the algebraic equations (\ref{oeq1}) and (\ref{pe1}). 
Then, using Eq. (\ref{secs}) we obtain the generalized susceptibility
\begin{equation}\label{suc}
\chi({\bf k},\omega)=
\chi_e ({\bf k}, \omega )
+\frac{\tilde{g}_{j}^2\varepsilon_e({\bf k},\omega)/NM'
-4\pi \tilde g_E\bar g_E(\tilde{\omega}^2_{j}-\omega^2)
-4\pi\tilde g_{j}(\tilde g_E \widetilde {Z}+\bar {g}_E \bar {Z})/M'}
{(\tilde{\omega}^2_{j}-\omega^2)
\varepsilon_e({\bf k},\omega)+4\pi N\widetilde{Z}\bar{Z}/M'}.
\end{equation}

The expression (\ref{suc}) is our main result. Here the poles of the second 
term give the spectrum of  collective excitations of the 
electron-phonon system. We discuss Eq. (\ref{suc}) in 
 various limiting cases.

\subsection*{ The  electronic scattering } 

We obtain the Raman electronic scattering from Eq. (\ref{suc}),
if we set $\tilde g_j=g_E=\widetilde Z= \bar Z =0$. We get
\begin{equation}\label{ers}
\chi({\bf k},\omega)=
\chi_e ({\bf k}, \omega )
+\frac{4\pi \tilde g_E^2}{\varepsilon_e({\bf k},\omega)},
\end{equation}
where $\tilde g_E$ is given in Eq. (\ref{oege}) with $g_E=0$.

For the isotropic Fermi surface, we  calculate the dielectric
function Eq. (\ref{dfu1}) performing  the integration:
\begin{equation}\label{del}
\langle 1/\Delta_p(k)\rangle=
\frac{1}{2 kv_F}\ln{\frac{1+\kappa}
{1-\kappa}},\quad \kappa=kv_F/(\omega+i\tau^{-1}),
\end{equation}
where one should  take the branch  of $\ln{x}$, 
which is  real for  positive real values of $x$.

For the anisotropic Fermi surface, the calculations can be done 
in  limiting cases.
For $|\kappa|\gg 1$, we use the expansion
 for the electronic dielectric function (\ref{dfu1}): 
\begin{equation} \label{as1}
\varepsilon_e({\bf k},\omega)=\varepsilon_{\infty}\{1+ (k_0/k)^2
[1+i(\pi\nu_0\omega/k)\langle\delta(\mu)/v\rangle]\},
\end{equation}
where $\mu={\bf v\cdot k}/vk$ and $\delta(x)$ is the Dirac
delta-function.
In this case, the Raman efficiency has a "tail"
due to the Landau damping:
\begin{equation}\label{ers1}
{\rm Im}\,\chi({\bf k}, \omega )=(\pi\nu_0\omega/k)
\langle \gamma^2({\bf p})\delta(\mu)/v\rangle.
\end{equation}
 We see, that the Raman cross section vanishes for
the isotropic vertex $\gamma({\bf p})$  because of Eq. (\ref{scr}). 
This is a result of 
the Coulomb screening. It was obtained first in the work \cite{AG}
for the Raman scattering in semiconductors (see, \cite{AF}).

In the  opposite case $|\kappa|\ll 1$, the first term in 
Eq. (\ref{ers}) gives a result
\begin{equation}\label{lw}
{\rm Im}\,\chi_e ({\bf k}, \omega )=\nu_0
\langle \gamma^2({\bf p})\rangle
\frac{\omega\tau}{(\omega\tau)^2+1},
\end{equation}
 which was found  first  in the work \cite{CZ} with a help
of the Green's function technique.
The second term in Eq. (\ref{ers}) reveals a plasmon pole at small values of
$k$.
The $k-$expansion of the dielectric function reads
\begin{equation}\label{dief}
\varepsilon_e(k,\omega)= \varepsilon_{\infty}
\left(1
-\frac{\omega_{pe}^2+k^2w}{\omega(\omega+i\tau^{-1})}\right),
\end{equation}
where the $k$-independent term represents the Drude conductivity and
the electron plasma frequency  is given by the integral over the Fermi
surface
$\omega_{pe}^2=k_0^2\langle v_z^2\rangle$.
The complex coefficient $w=k^2_0(\langle v_z^4\rangle
+i\langle v_z^2\rangle^2/\omega\tau)/(\omega+i\tau^{-1})^2$.
For the isotropic electron spectrum and in  collisionless regime, 
$\tau^{-1}=0$, the coefficient $w=(3/5)\omega_{pe}^2 v_F^2$.

The $k-$expansion of $ \tilde{g}_E$  gives
$$\tilde{g}_E = -ie\nu_0k\langle\gamma({\bf p})v_z^2\rangle/
(\omega+i\tau^{-1})^2,   $$
because $g_E=0$ and the zero-order term in the $k-$expansion
vanishes due to the time invariance ${\bf v}\rightarrow -{\bf v}$.
Then the intensity of the plasmon peak $\sim k^2$  
in accordance with the known
 behavior of the dynamical structure factor.

\subsection*{The  Raman scattering by  TO phonons} 

Ihe second term in
Eq. (\ref{suc}) gives the TO-phonon scattering, 
if we set $\widetilde{Z}=\bar Z= \tilde g_E=0$: 
\begin{equation}\label{suct}
\chi({\bf k},\omega)=
\frac{\tilde{g}_{\rm TO}^2/NM'}
{\tilde{\omega}^2_{\rm TO}-\omega^2-i\omega\Gamma^{\rm nat}},
\end{equation}
where $\tilde{\omega}_{\rm TO}$, $\tilde{g}_{\rm TO}$ are 
defined in Eqs. (\ref{shd}) and (\ref{doc}) with $\zeta_j({\bf p})=
\zeta_{\rm TO}({\bf p})$; we add the phonon  width $\Gamma^{\rm nat}$ 
mentioned above.

Here we should note two points. First, the TO-resonance takes place
at the renormalized frequency $\tilde{\omega}_{\rm TO}$. Taking the real and 
imaginary parts of
the expression  (\ref{shd}), we obtain the
TO-phonon shift and width due to the deformation interaction $\zeta({\bf p})$
with carriers:
$$ \Delta\omega_k={\rm Re}\,(\tilde{\omega}^2_{\rm TO}-\omega^2_k)/2\omega_k,
\quad \Gamma=\Gamma^{\rm nat}-{\rm Im}\,\tilde{\omega}^2_{\rm TO}/\omega_k.$$
Second, because of the interaction 
 with carriers, the coupling $\tilde{g}_{\rm TO}$  (\ref{doc}) has 
 an imaginary part.
Therefore the line-shape of the resonance becomes asymmetric 
(the Fano resonance):
\begin{equation}\label{suct1} 
{\rm Im}\,\chi ({\bf k}, \omega )=\frac{1}{NM'}\frac{\omega\Gamma g_{\rm TO}^2+
[({\rm Re}\,\tilde{\omega}_{\rm TO})^2-\omega^2]{\rm Im}\,\tilde g_{\rm TO}^2}
{[({\rm Re}\,\tilde{\omega}_{\rm TO})^2-\omega^2]^2+(\omega\Gamma)^2}.
\end{equation}
The line-shape asymmetry depends 
on the sign of
 ${\rm Re}\, g_{\rm TO}$. For instance, if ${\rm Re}\, g_{\rm TO}>0$, 
 the high-frequency
 wing of the resonance line drops more slowly than the low-frequency one.
In the limiting case $  \kappa\gg 1$, 
we  expand 
\begin{equation}\label{rgso}
\tilde{g}_{\rm TO}= g_{\rm TO}+
(\nu_0\omega/k)\langle\gamma({\bf p}) \zeta_{\rm TO}({\bf p})
(-i\pi+2\omega/kv)\delta(\mu)/v\rangle
\end{equation}
and for $ \kappa\ll 1$:
\begin{equation}\label{rgto}
\tilde{g}_{\rm TO}=g_{\rm TO}+
\frac{\omega\nu_0}{\omega+i\tau^{-1}}
\left(\langle\gamma({\bf p}) \zeta_{\rm TO}({\bf p})\rangle+
\frac{k^2\langle v_z^2\gamma ({\bf p})\zeta_{\rm TO}({\bf p})\rangle}
{(\omega+i\tau^{-1})^2}\right).
\end{equation}
  
Notice, that  the electron-phonon interaction  $\zeta_{\rm TO}({\bf p})$ 
and the light scattering  $\gamma ({\bf p})$ by carriers renormalize jointly 
the coupling $g_{\rm TO}$.
The frequency renormalization 
$\tilde{\omega}^2_{\rm TO}$ [see Eq. (\ref{shd})] results only from
the electron-phonon interaction  $\zeta_{\rm TO}({\bf p})$. 
The corresponding expressions  can be
obtained from Eqs. (\ref{rgso}) and (\ref{rgto}) by substitution
$\gamma({\bf p})\rightarrow\zeta_{\rm TO}({\bf p})$.
 We see, that   the TO phonons become broader and harder
because of the interaction with carriers. 
 
 Emphasize, that the phonon renormalizations  depend 
 on the carrier density $\nu_0$
and  the average coupling $\zeta_j({\bf p})-\langle\zeta_j({\bf p})\rangle$. 
They vanish for the
isotropic Fermi surface. The  maximum  value
of the relative renormalization has the order of 
$\lambda ap_Fm^{*}\omega/m|\omega+i\tau^{-1}|$ at 
$ kv\sim|\omega+i\tau^{-1}|$, where $\lambda$ is the dimensionless
electron-phonon coupling and $m^{*}$ is effective electron mass.
 
\subsection* { The Raman scattering by  LO-phonon--plasmon coupled modes} 

In this case, the  carriers interact with each other and 
the ion vibrations 
 via both the macroscopic electric field $E({\bf r}, t)$ and
 the deformation potential $\zeta_{\rm LO}({\bf p})$. 
In the long-wave limit $k\rightarrow 0$,
 Eqs.
(\ref{oeq2d}), (\ref{oeq3d}),  and (\ref{oege}) show 
no renormalization of the ionic charge, 
 $\widetilde Z=\bar Z=Z$, and 
 the electro-optic constant, $\tilde g_E=g_E$. The equation (\ref{suc}) 
 takes the form:
 \begin{equation}\label{suc2}
\chi(0,\omega)=
\chi_e (0, \omega )
+\frac{\tilde{g}_{\rm LO}^2\varepsilon_e(0,\omega)/NM'
-4\pi  g_E^2(\tilde{\omega}^2_0-i\omega\Gamma^{\rm nat}-\omega^2)-
8\pi g_E \tilde{g}_{\rm LO}Z/M'}
{(\tilde{\omega}^2_0-i\omega\Gamma^{\rm nat}-\omega^2)
\varepsilon_e(0,\omega)+4\pi NZ^2/M'},
\end{equation}
where the first term is given in Eq. (\ref{lw}).
 The deformation potential $\zeta_{\rm LO}({\bf p})$ renormalizes  
 the phonon frequency
$\tilde\omega_0$  (\ref{shd}), as well as  
the deformation-optic constant $\tilde{g}_{\rm LO}$  (\ref{doc}).
The corresponding expansion in the limiting cases
 are similar to  Eqs. (\ref{rgso}) and
(\ref{rgto}). All mentioned above about the TO line asymmetry concerns
also the LO line.

Because the dielectric function of the electron-ion system  reads
\begin{equation}\label{dif}
\varepsilon( 0,\omega)=
\varepsilon_e( 0,\omega)+4\pi
NZ^2/M'(\tilde{\omega}^2_0-i\omega\Gamma^{\rm nat}-\omega^2),
\end{equation}
the second term on the right-hand side in Eq. (\ref{suc2}) 
has  poles  while $\varepsilon( 0,\omega)=0$.
This condition defines the frequency of 
coupled phonon-plasmon modes in the long-wave
limit.

In the absence of the electron and phonon collisions
($\tau^{-1}=\Gamma^{\rm{nat}}=0$), and without the el-ph interaction
($\zeta({\bf p})=0$),
one obtains with the help of Eq. (\ref{dief})  
the biquadratic equation. It gives
 the frequencies of the coupled phonon-plasmon modes at
$k=0$:
\begin{equation} \label{pm}
\omega^2_{\pm}=
\frac{1}{2}
(\omega_{pe}^2+\omega_{\rm{LO}}^2)
\pm\frac{1}{2}\left[
(\omega_{pe}^2+\omega_{\rm{LO}}^2)^2-4\omega_{pe}^2\omega_{\rm{TO}}^2
\right]^{1/2},
\end{equation}
where $\omega_{\rm{TO}}=\omega_k$ is the TO-mode frequency at $k=0$,
$\omega_{\rm{LO}}^2=\omega_{\rm{TO}}^2+\omega^2_{pi}$, and 
$\omega^2_{pi}=4\pi NZ^2/\varepsilon_{\infty}M'$.
These frequencies (related to the $\omega_{\rm{TO}}$) are shown
in Fig. 1  as functions of the electron concentration,
 namely, $\omega_{pe}/\omega_{\rm{TO}}$.
The upper  line begins at
$\omega_{\rm{LO}}$ and tends to the electron plasma frequency
$\omega_{pe}$.
The lower frequency  starts as
$\omega_{pe}\omega_{\rm{TO}}/\omega_{\rm{LO}}$ and then approaches
 $\omega_{\rm{TO}}$.
In other words,
observing in the optic range the longitudinal phonon mode and
adding electrons,
we see a transition of the longitudinal phonon frequency
from
$\omega_{\rm{LO}}$  to $\omega_{\rm{TO}}$. This is a result of
the Coulomb screening. 

We can compare Eq. (\ref{suc2}) with the theory of Hon and Faust \cite{HF}.
 Since in that theory the electron--phonon interaction 
 $\zeta_{\rm LO}({\bf p})$ as well as the electronic scattering 
 $\gamma({\bf p})$
 were ignored, the phonon frequency  and the 
 deformation-optic constant were not renormalized. Then Eq. (\ref{suc2})
 can be rewritten as 
 \begin{equation}\label{suc3}
\chi( 0,\omega)=
\frac{(4\pi g_{E})^2}{\varepsilon_{\infty}\varepsilon(0,\omega)}
[\varepsilon_e(0,\omega)A^2\chi_{I}/\varepsilon_{\infty}-
\varepsilon_{\infty}/4\pi-2A\chi_I],
\end{equation}
where
$\chi_{I}= NZ^2/M'(\omega_{\rm TO}^2-i\omega\Gamma^{\rm nat}-\omega^2),$
$A= C\omega_{\rm TO}^2 M'\varepsilon_{\infty}/4\pi NZ^2,$
and $C=g_{\rm LO}Z/g_E M\omega^2_{\rm TO}$ is the Faust-Henry coefficient.
Now we see, that the expression (\ref{suc3}) 
coincides with the result of Hon and Faust
[see, e.g. the paper \cite{ACI}, Eq. (3.1) ].

For $k\ne 0 $, Eq. (\ref{suc}) includes the dielectric function (\ref{dfu1})
which differs from the Lindhard-Mermin expression.
 Second, the condition determining  frequencies and  damping 
 of the phonon-plasmon coupled modes
\begin{equation}\label{dise}
{(\tilde{\omega}^2_{j}-i\omega\Gamma^{\rm nat}-\omega^2)
\varepsilon_e({\bf k},\omega)+4\pi N\widetilde{Z}\bar{Z}/M'}=0
\end{equation}
contains the phonon frequency $\tilde\omega$ and the
ionic charge  renormalized by the electron-phonon interaction
$\zeta_{\rm LO}({\bf p})$. Third,
the electro-optic
coupling  $g_E$ (\ref{oege}) is modified  due to the light scattering 
$\gamma({\bf p})$ from the electron-hole pairs. This effect is not canceled 
in the product 
$\tilde{g}_E\bar{g}_E$ of Eq. (\ref{suc})   even in the absence of
the electron-phonon interaction $\zeta_{\rm LO}({\bf p})$. The expansion of 
$\tilde{g}_E$ has the form 
\begin{equation}\label{gk}
\tilde{g}_E=g_E-\frac{ie\nu_0k\langle
v_z^2\gamma({\bf p})\rangle}{(\omega+i\tau^{-1})^2}
\end{equation}
for $ |\kappa|\ll 1$ and 
\begin{equation}\label{gk1}
\tilde{g}_E=g_E+
e\nu_0k^{-2}(\omega+i\tau^{-1})\langle\gamma({\bf p}) 
(-\pi/2-i\omega/kv)\delta(\mu)/v\rangle
\end{equation}
for $ |\kappa|\gg 1$.
We  note, that the term $\tilde{g}_E\bar{g}_E$ has the largest
imaginary part for $\omega\tau \simeq 1$ and then  results more 
in the line-shape  asymmetry.

Schematicly, the dispersion of phonon-plasmon modes is shown in Fig. 2.
There are two main specularity of this figure. First, the behavior
of the upper mode near the line $\omega=kv_F$. Around this line
($\tau^{-1}<\omega-kv_F<<kv_F$)
the dielectric function (\ref{dfu1}) has a singularity
\begin{equation} \label{eps1}
\varepsilon_e(k,\omega)= \varepsilon_{\infty}+
\varepsilon_{\infty}\frac{k_0^2}{k^2}\left\{
1-\frac{\omega}{2 kv_F}\left[\frac{1}{2}\ln{\frac{4k^2v_F^2}
{(\omega-kv_F)^2+\tau^{-2}}}-i\frac{\tau^{-1}}{\omega-kv_F}\right]\right\}.
\end{equation}
Because of this singularity, the upper mode approaches the asymptote $\omega=kv_F$,
while the wave vector $k$ increases.
Second, in the region $ kv_F\gg \omega $, there is always one mode which mainly has 
phonon character. The reason of that is the decrease with $k$ of the imaginary part
of the dielectric function  (\ref{as1}).

\section{Discussion} 

Now we consider obtained results in a simplest way. Let us assume  
that the electronic scattering is negligibly 
small, $\gamma({\bf p})=0$. The second term in the parentheses
of Eq. (\ref{shd}) is less than the first one in both limiting cases
$\kappa\ll 1$ and $\kappa\gg 1$. We neglect this term at all.  
We also do not take into account     the ionic charge renormalization
because it vanishes at  small values of $\kappa$. Then we can use the expression 
(\ref{del}) not only for the dielectric function $\varepsilon_e({\bf k},\omega)$,
Eq. (\ref{dfu1}), but also for the renormalized phonon frequency  
$\tilde{\omega}$,
Eq. (\ref{shd}). 

Results of the  numerical calculations of the
Raman spectra, Eq. (\ref{suc}), in this approximation
are shown in Figs. 3 and 4 for two values of the electron-phonon 
coupling 
$\lambda_{eph}=\nu_0\langle\zeta^2({\bf p})\rangle/\omega_0^3M'N\simeq p_Fam^*/m.$
We take  values of the Faust-Henry coefficient $C = - 0.5$ and the
phonon natural width $\Gamma^{\rm nat}= 10^{-2}\omega_{\rm TO}$. 
The electron collision rate is taken as $\tau^{-1}= 10^{-1}\omega_{\rm TO}$, 
which is the usual value for heavily doped semiconductors \cite{ACI},\cite{FP}.
The wave vector $k$ and the Thomas-Fermi parameter $k_0$ are given in 
units of $\omega_{\rm TO}/v_F$, and the frequency $\omega$ in units of
$\omega_{\rm TO}$ in all figures. Both these figures correspond to the case of 
small carrier numbers $\omega_{pe}<\omega_{\rm TO}$ (see, the right panel
of Fig. 2; for the quadratic electron spectrum $\omega_{pe}=k_0v_F/\sqrt{3}$). 
The left peak mainly has a plasmon character  and the right peak is mainly
the LO phonon. We put  the ion plasmon frequency 
$\omega_{pi}=\omega_{\rm TO}$, therefore $\omega_{\rm LO}=
\sqrt{2}\omega_{\rm TO}$. When the wave vector $k$ approaches the boundary of
the Landau damping region $kv_F> \omega$, the plasmon peak becomes broader
and almost disappears    at $k=0.8$.  The broad continuum in the region  
$kv_F> \omega$ results from the excitation of electron-hole pairs. The
intensity of the plasmon peak becomes larger in the comparison to the phonon
peak, when the electron-phonon interaction $\lambda_{eph}$ increases.

The $k-$dispersion of the plasmon  
(the peak position   of the Raman spectra as a function of $k$), 
the line width (the full width at half maximum), and the line asymmetry 
(the frequency difference between the right and 
left wings of the resonance line at half maximum)  
are shown in Figs. 5 - 6, all in units of $\omega_{rm TO}$. 
The width and asymmetry become much larger while
the plasmon peak is immersed in the electron-hole continuum. 
The maximum in this region of spectra is nothing  but
 the electron-hole contribution mainly. In Fig. 6, we see for 
 $\lambda_{eph}=0.1$ how close this maximum is located 
 to the line $\omega=kv_F$.  

The behavior of the phonon peak around $\omega=\omega_{\rm LO}$ 
with increasing of $k$ is shown in Fig. 7. While the wave vector increases
from $k=0$ to $k=1.7$, the phonon peak is evidently 
shifted to the higher frequency and becomes broader. This is effect of
the Landau damping (see Fig. 2, the right panel). But then, $k>1.75$,
this peak appears at the lower frequency, $\omega\simeq 1.4$ and becomes
more sharp for $k>2.2$, because the Landau damping decreases with $k$
[see Eq. (\ref{as1})].

The Raman spectra for  heavily doped semiconductors and metals are
shown in Figs. 8 - 9 (see also the left panel of Fig. 2). The phonon peak
is located now around $\omega\sim \omega_{\rm TO}$ instead of
$\omega\sim \omega_{\rm LO}$. This is effect of the Coulomb screening:  carriers
decrease the frequency of the LO mode from $\omega_{\rm LO}$ 
to $\omega_{\rm TO}$. We see also that the el-ph interaction 
suppresses the phonon peak. 

The effect of the Coulomb screening and 
the el-ph interaction on
the phonon mode is  clearly seen in Figs. 10 - 11, where the phonon part
of spectra is shown in detail. The lines are
very asymmetric. The phonon dispersion, the line
width, and the line asymmetry as  $k-$functions are shown in Fig. 12. 
 We see a singularity
at $k\simeq \omega/v_F$. It is interesting to estimate the value of
the phonon dispersion. With the help of Fig. 11, we wind 
$d\omega/dk\le 10^{-1} v_F$. From the other side, using Eqs. (\ref{as1}),
(\ref{dief}), we find for the
phonon dispersion $\omega^2=\omega^2_{\rm TO}+\omega^2_{pi}k^2/k_0^2$,
which well corresponds  with the previous estimation for our values of 
$k_0$ and $\omega_{pi}$. Note, that these estimations  confirm the
adiabatic approximation, since the value of dispersion
$s=\omega_{pi}/k_0\sim v_F\sqrt{m/M}$ contains the  adiabatic parameter.

 In Fig. 13, the  dispersion, the line width, and the line
 asymmetry are shown for the plasmon peak in heavily doped semiconductors.
 Here the effect of the el-ph interaction on the phonon dispersion is weak, no
 influence on the width and asymmetry of the line is seen.

\section{Conclusions}

In conclusions, we emphasize first, that  the result of the paper (\ref{suc})
describes the renormalization of the phonon frequencies, the effective
ionic charge and
the coupling constants due to the electron--phonon  deformation interaction
$\zeta_j({\bf p})$. Second,  it involves the $k-$dependent semiclassical dielectric
function instead of the Lindhard-Mermin expression. At last,
 the light scattering $\gamma({\bf p})$ with excitations of the electron-hole pairs 
  gives not only an additional contribution $\chi_e$ in Eq. (\ref{suc}), 
  but also modifies
  the electro-optic $g_E$ and deformation-optic $g_j$ coupling constants
  which become dependent on the frequency and momentum transfers.

The author acknowledges the kind hospitality of the Max-Plank-Institut f\"{u}r
Physik komplexer Systeme, Dresden, were this work was completed.
The work was partially supported by the RFBR
(project 01-02-16211).

\clearpage
\begin{figure}
\begin{center}
\epsfxsize=70mm
\epsfysize=100mm
\epsfbox{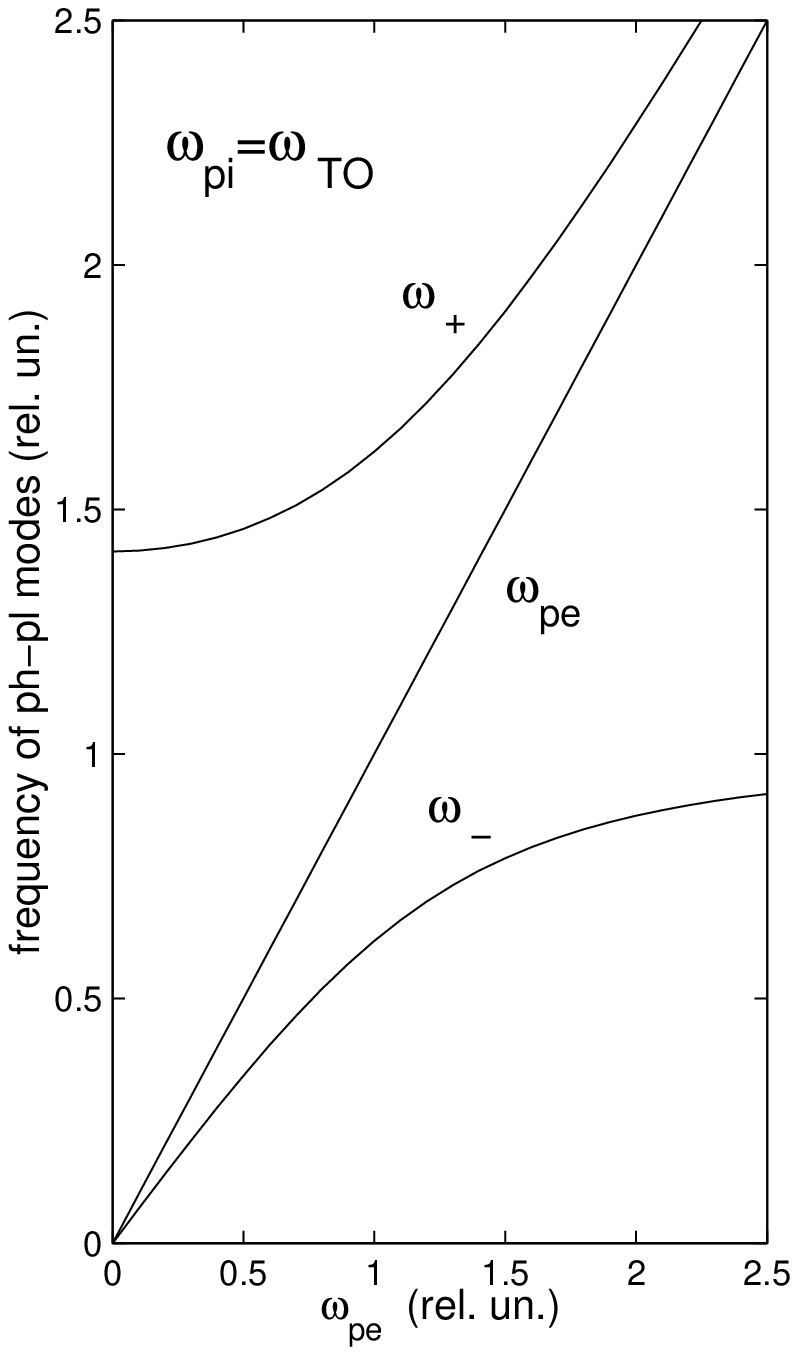}
\end{center}
\caption{Frequencies (in units of $\omega_{\rm{TO}}$)
of the phonon-plasmon modes at $k=0$
in dependence of the free carrier concentration, namely, of
 the electron plasma frequency
(in units of $\omega_{\rm{TO}}$). We set the ion plasma frequency
$\omega_{pi}=\omega_{\rm{TO}}$ in the absence of the free carriers.
Then $\omega_{\rm{LO}}/\omega_{\rm{TO}}=\sqrt{2}$. }
\end{figure}

\clearpage
\begin{figure}
\begin{center}
\epsfxsize=140mm
\epsfysize=110mm
\epsfbox{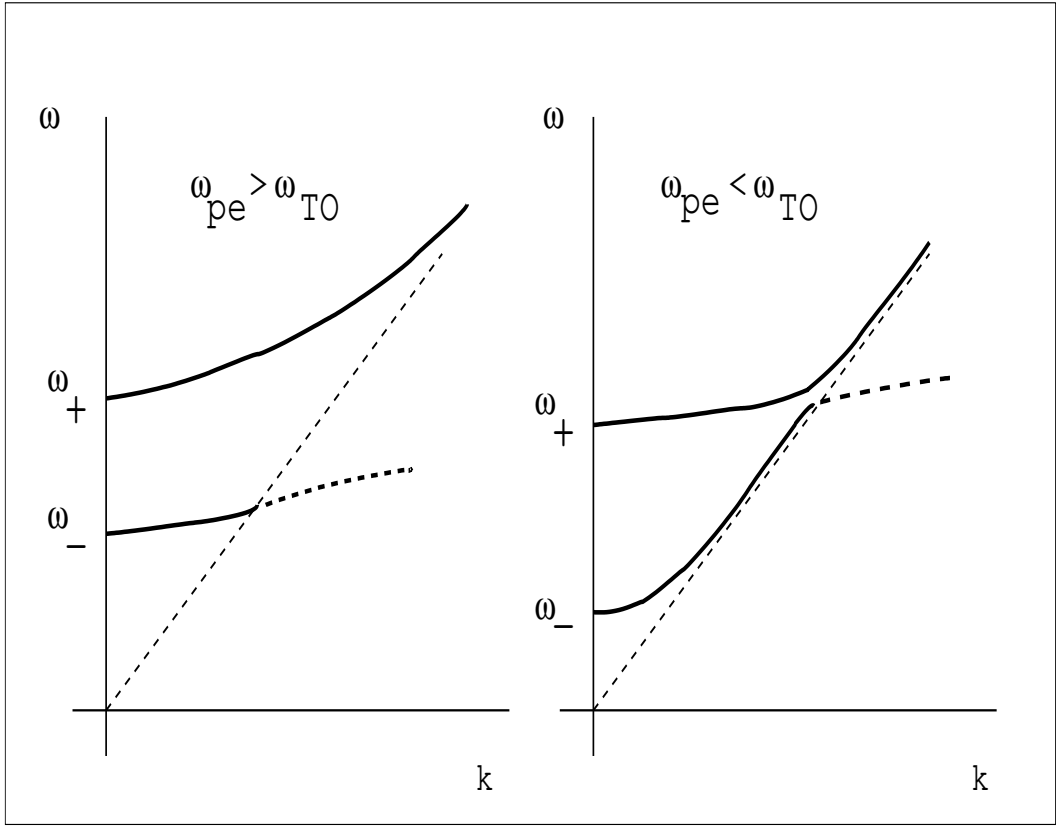}
\end{center}
\caption{Schematic drawing of the dispersion of phonon-plasmon 
modes for the metallic
($\omega_{pe}>\omega_{\rm{TO}}$, left panel) and
semiconducting ($\omega_{pe}<\omega_{\rm{TO}}$, right panel)
 carrier concentration.
The dashed straight lines separate the domain
$kv_F>\omega$, where the Landau damping exists; 
the dashed curves represent   damped modes there.  }
\end{figure}
 
\clearpage
\begin{figure}
\begin{center}
\epsfxsize=140mm
\epsfysize=110mm
\epsfbox{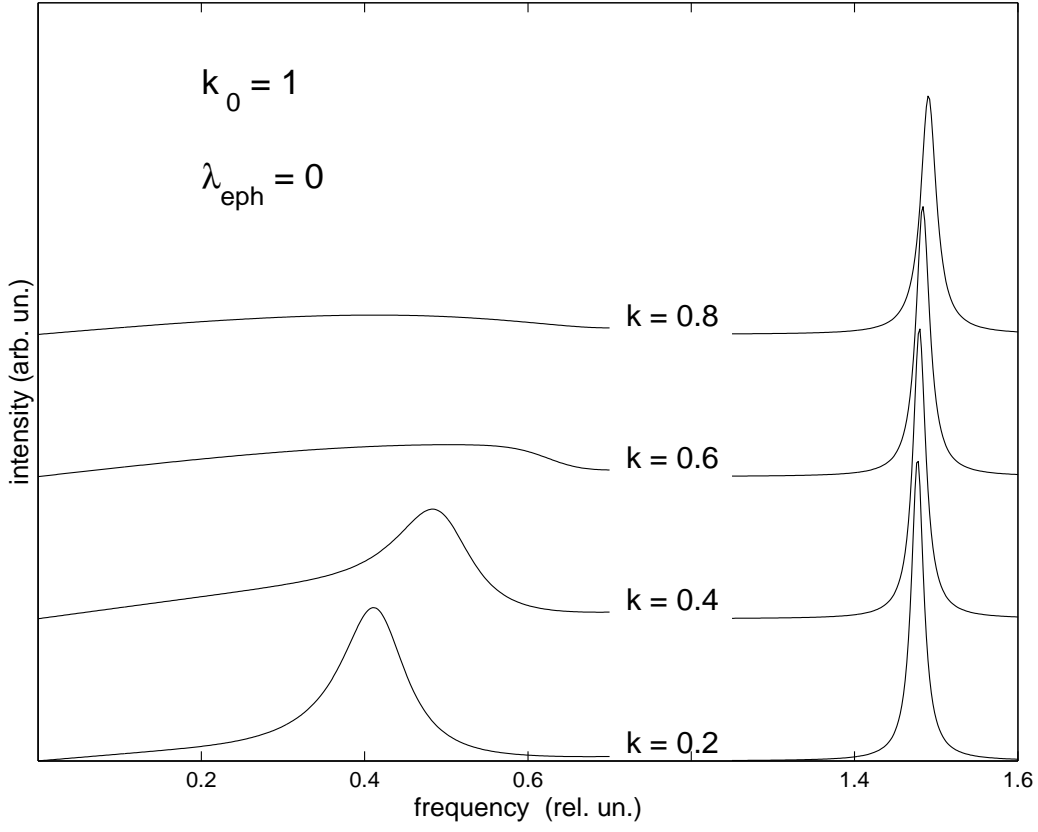}
\end{center}
\caption{ Raman spectra from a semiconductor with 
low carrier concentration
as a function of the frequency transfer $\omega$ for denoted values
of the momentum transfer $k$, the Thomas-Fermi
parameter $k_0$ (in  units of $\omega_{\rm{TO}}/v_F$),
and the electron-phonon coupling constant $\lambda_{eph} $. 
We set the ion plasma frequency
$\omega_{pi}=\omega_{\rm{TO}}$, the phonon natural width
$\Gamma^{\rm{nat}}/\omega_{\rm{TO}}=10^{-2}$,
and the carrier relaxation rate $\tau^{-1}/\Gamma^{\rm nat}=10$. }
\end{figure}

\clearpage
\begin{figure}
\begin{center}
\epsfxsize=140mm
\epsfysize=110mm
\epsfbox{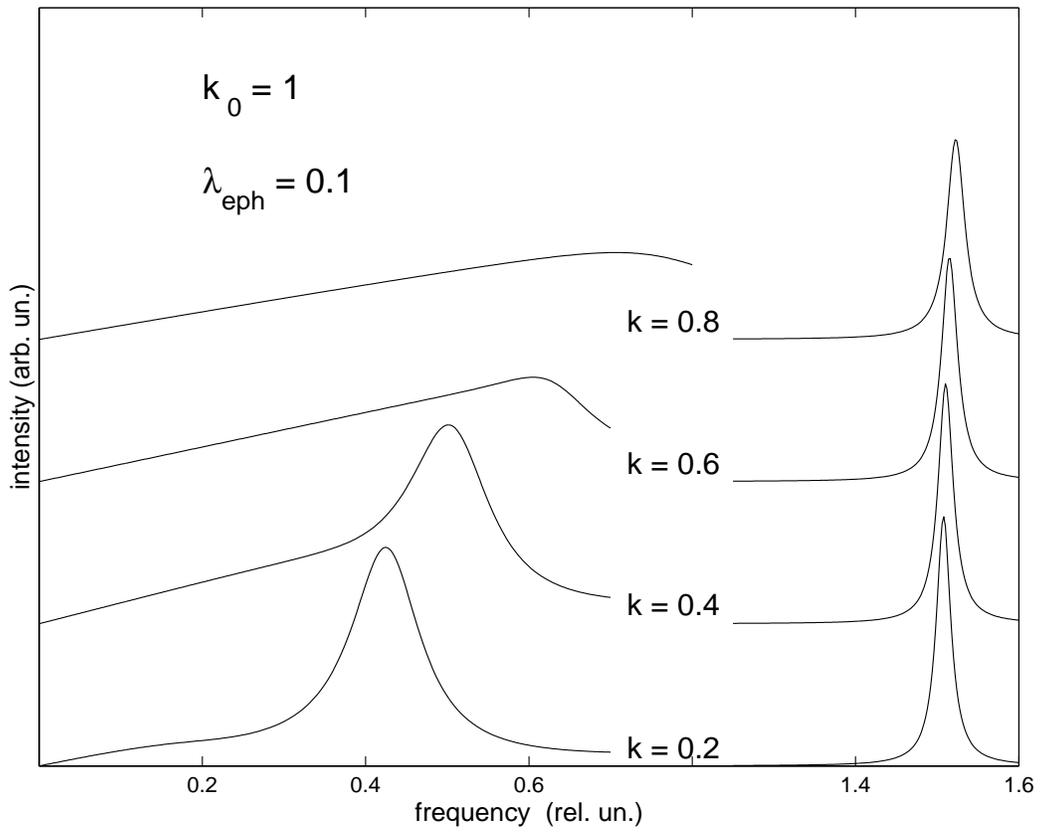}
\end{center}
\caption{Same as Fig. 3  for $\lambda_{eph}= 0.1$.}
\end{figure}

\clearpage
\begin{figure}
\begin{center}
\epsfxsize=140mm
\epsfysize=110mm
\epsfbox{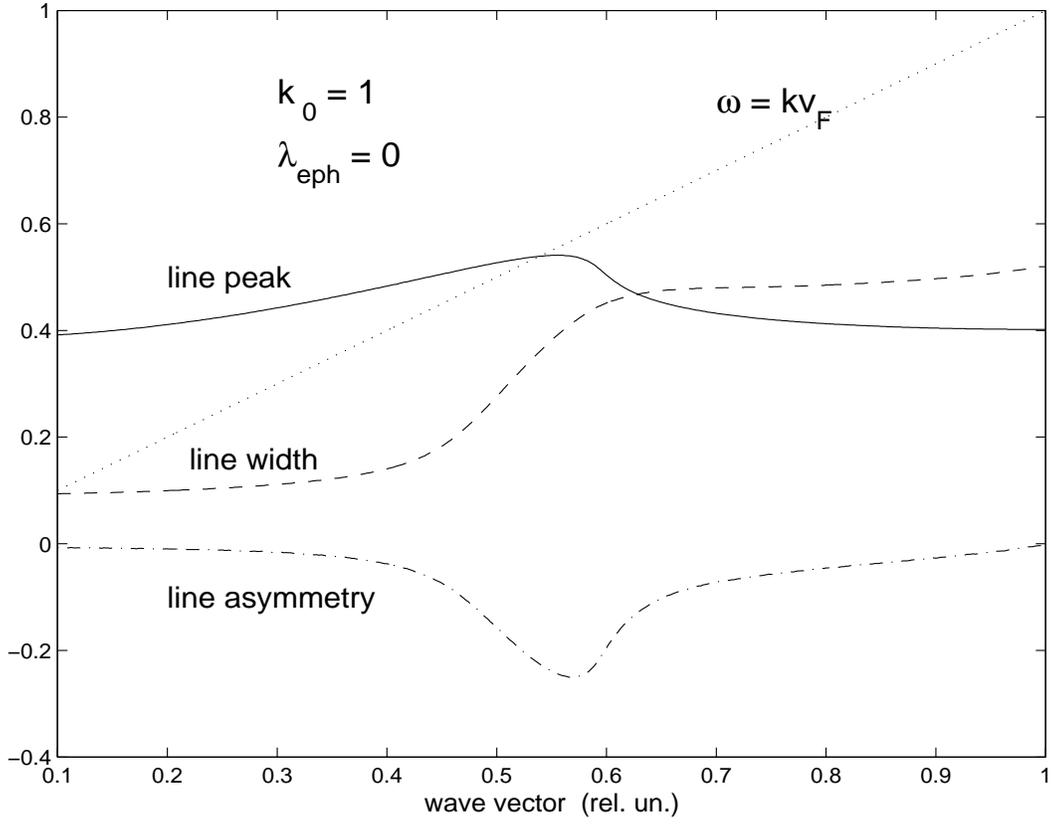}
\end{center}
\caption{The plasmon dispersion as a function of  $k$ in
units of $\omega_{\rm TO}/v_F$
(the position  in units of $\omega_{\rm TO}$ 
of the  line peak of  Raman spectra;  
upper part of the figure,  solid line). In the bottom, the line
width (the full width at half maximum,  dashed line), and the line asymmetry 
(the difference between the right and left wings at half maximum, 
dash-dotted line)
 in units of $\omega_{\rm TO}$. The Landau damping exists to the right of the
 dotted line $\omega=kv_F$.}
\end{figure}

\clearpage
\begin{figure}
\begin{center}
\epsfxsize=140mm
\epsfysize=110mm
\epsfbox{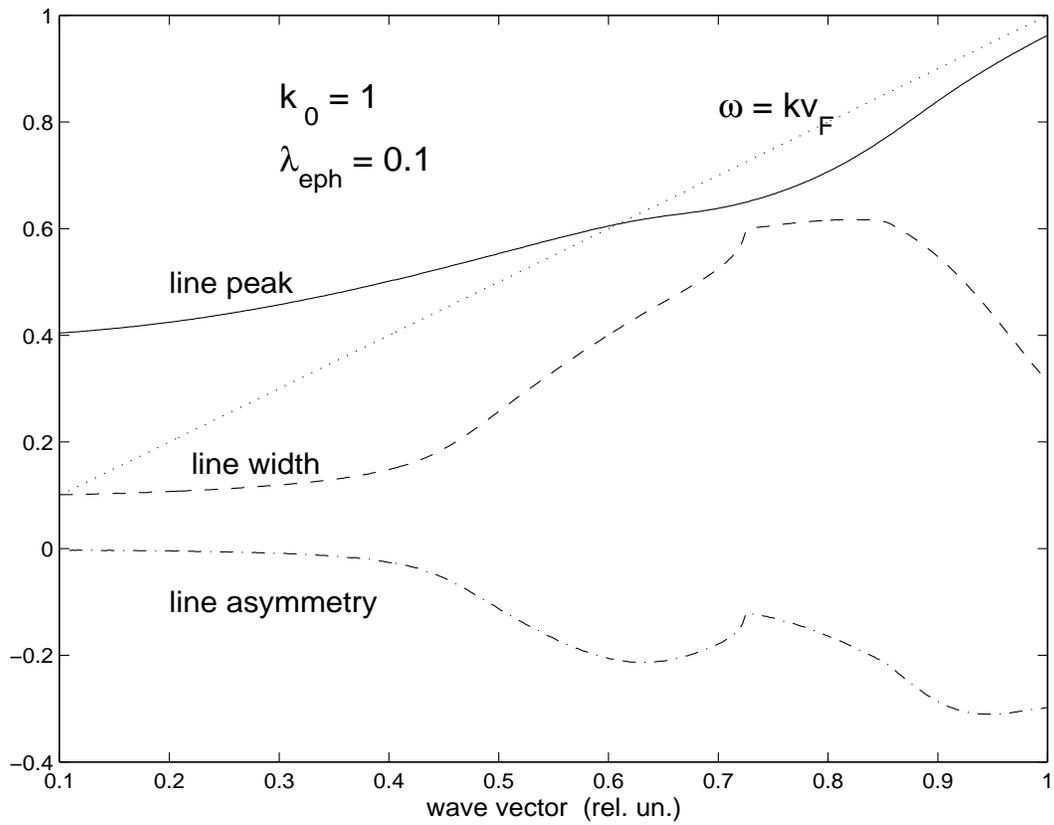}
\end{center}
\caption{Same as Fig. 5  for the electron-phonon coupling 
$\lambda_{eph}= 0.1$. }
\end{figure}

\clearpage
\begin{figure}
\begin{center}
\epsfxsize=140mm
\epsfysize=110mm
\epsfbox{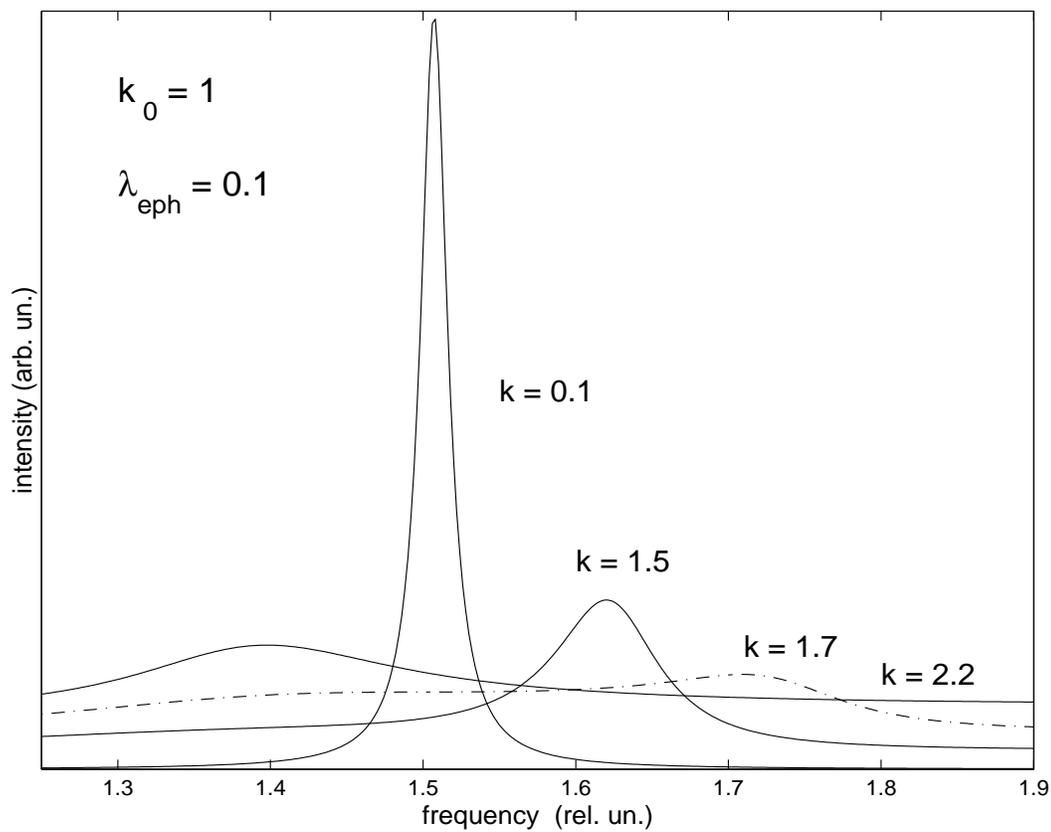}
\end{center}
\caption{ The LO phonon Raman spectra for  large momentum transfers $k$.}
\end{figure}

\clearpage
\begin{figure}
\begin{center}
\epsfxsize=140mm
\epsfysize=110mm
\epsfbox{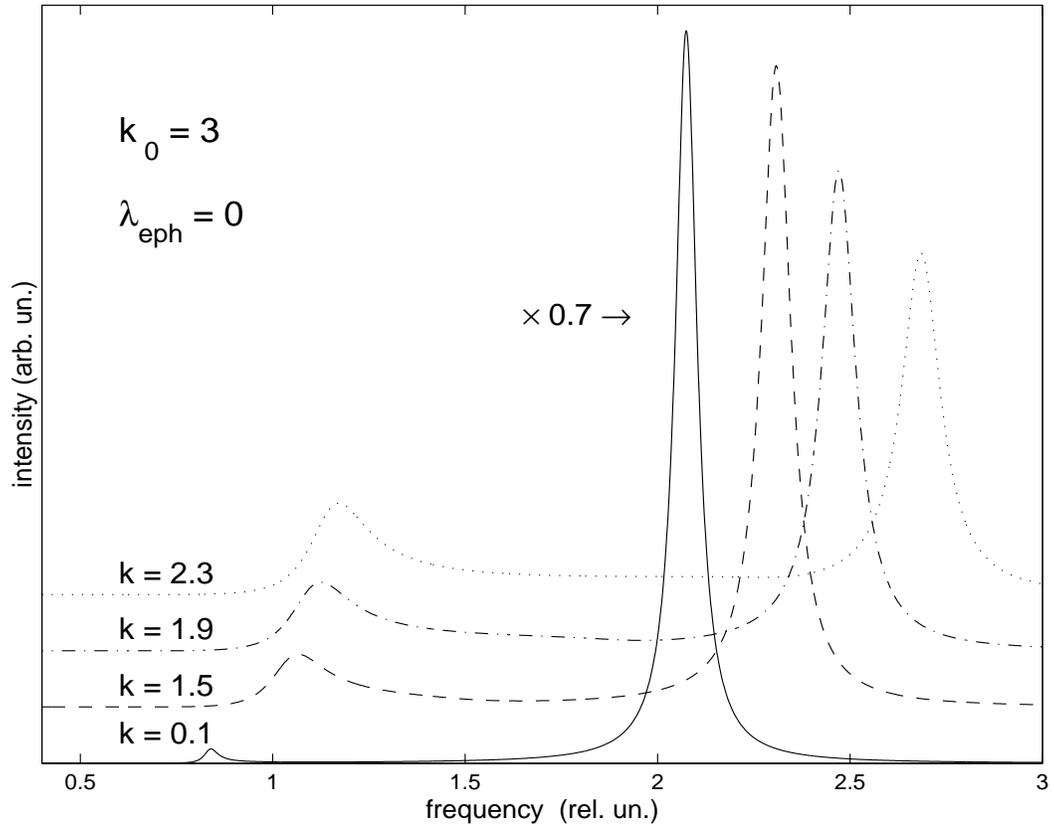}
\end{center}
\caption{Raman spectra from a heavily doped semiconductor; notations as
in Fig. 1.  }
\end{figure}

\clearpage
\begin{figure}
\begin{center}
\epsfxsize=140mm
\epsfysize=110mm
\epsfbox{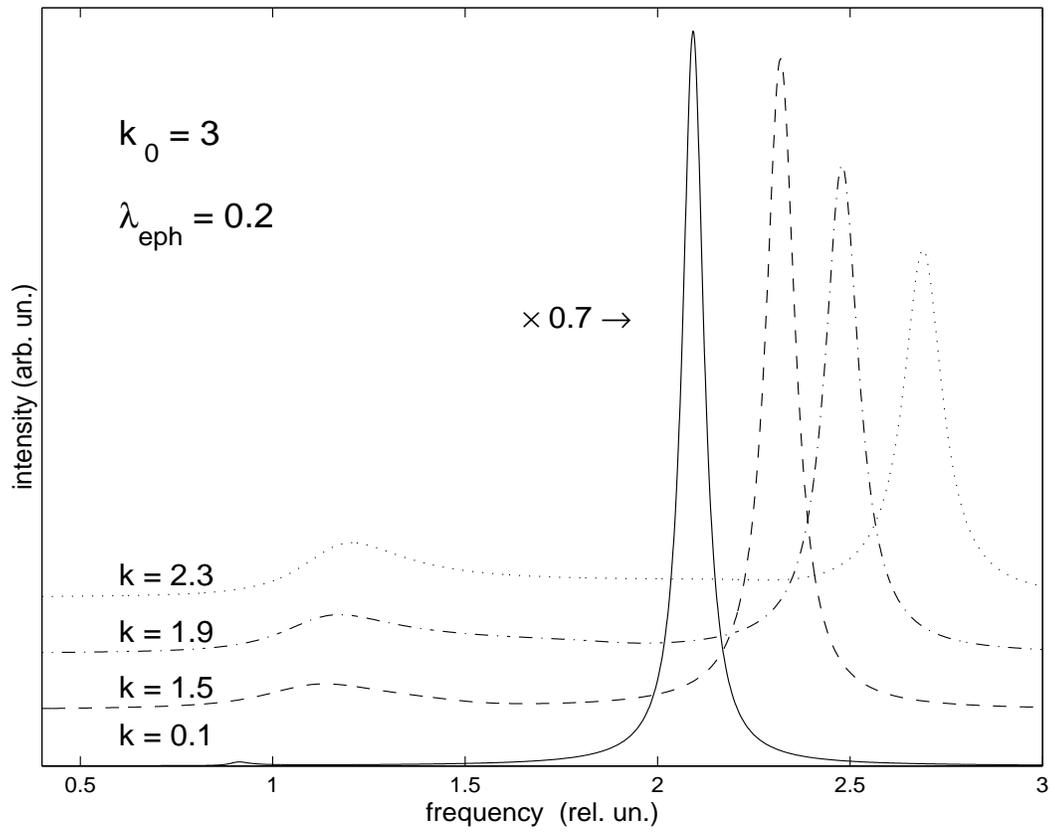}
\end{center}
\caption{Same as in Fig. 8  for $\lambda_{eph}=0.2$. }
\end{figure}

\clearpage
\begin{figure}
\begin{center}
\epsfxsize=140mm
\epsfysize=110mm
\epsfbox{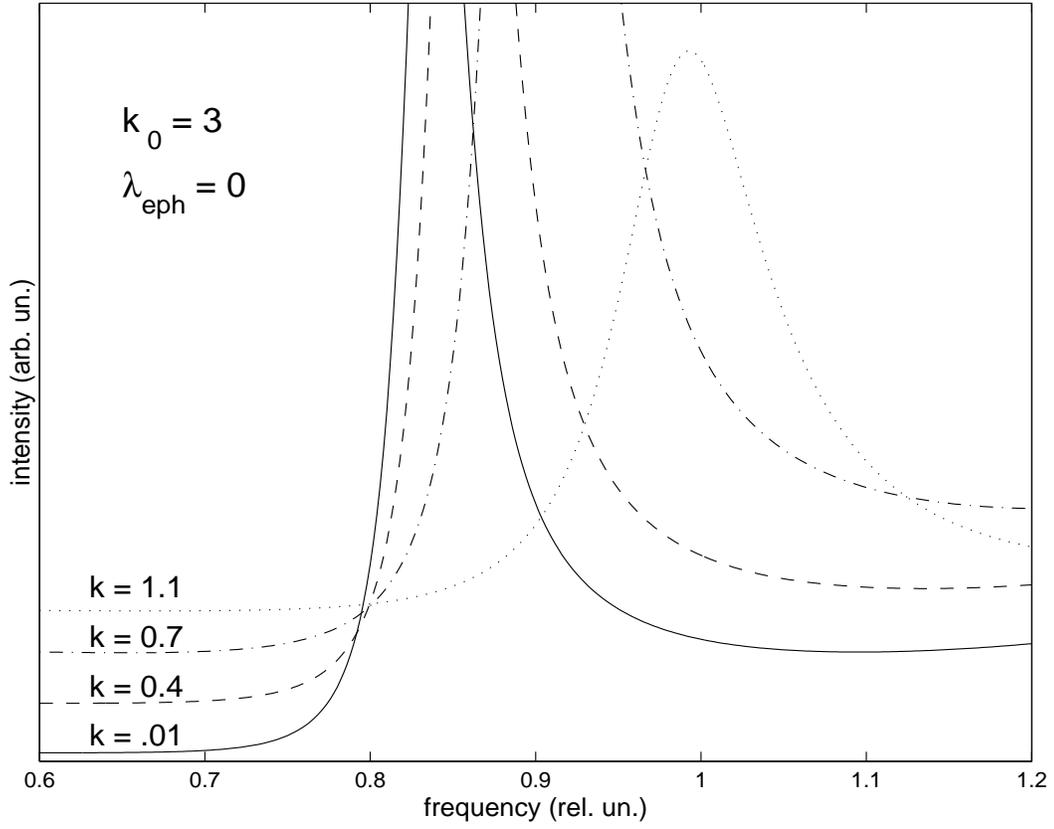}
\end{center}
\caption{The LO phonon part of the Raman spectra from heavily doped
semiconductor for  various momentum transfers. }
\end{figure}

\clearpage
\begin{figure}
\begin{center}
\epsfxsize=140mm
\epsfysize=110mm
\epsfbox{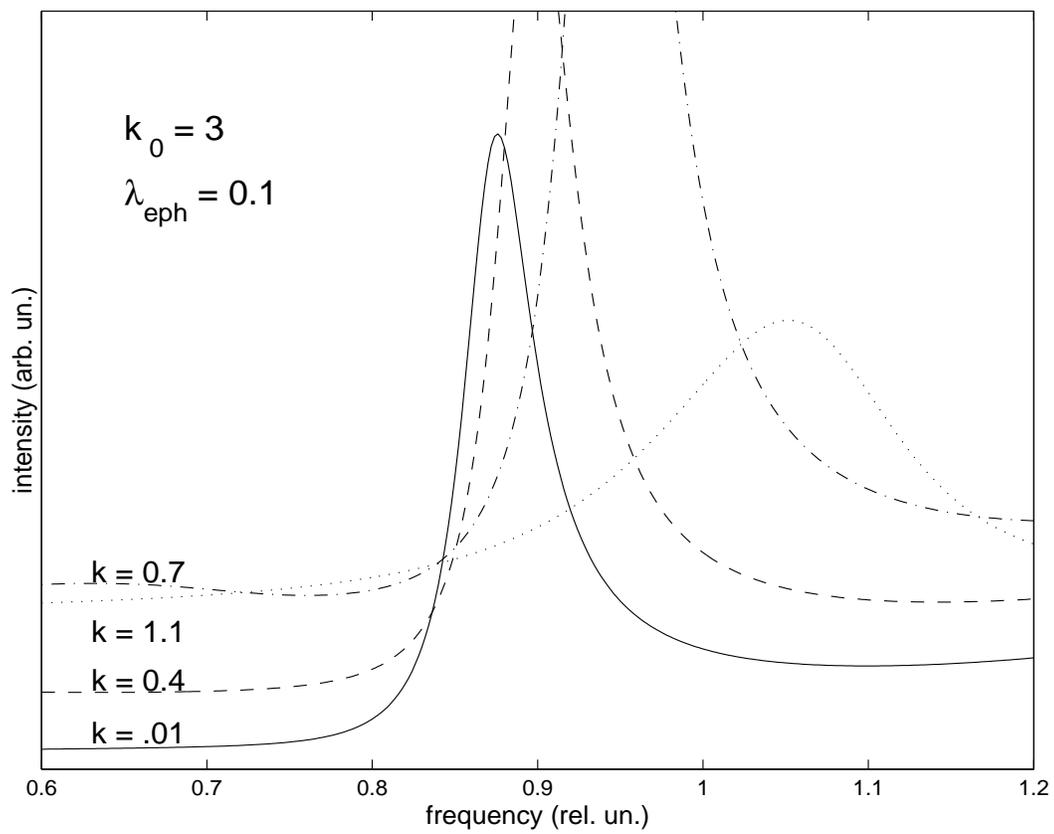}
\end{center}
\caption{Same as in Fig 10, but the el-ph interaction is taken into
account. }
\end{figure}

\clearpage
\begin{figure}
\begin{center}
\epsfxsize=140mm
\epsfysize=110mm
\epsfbox{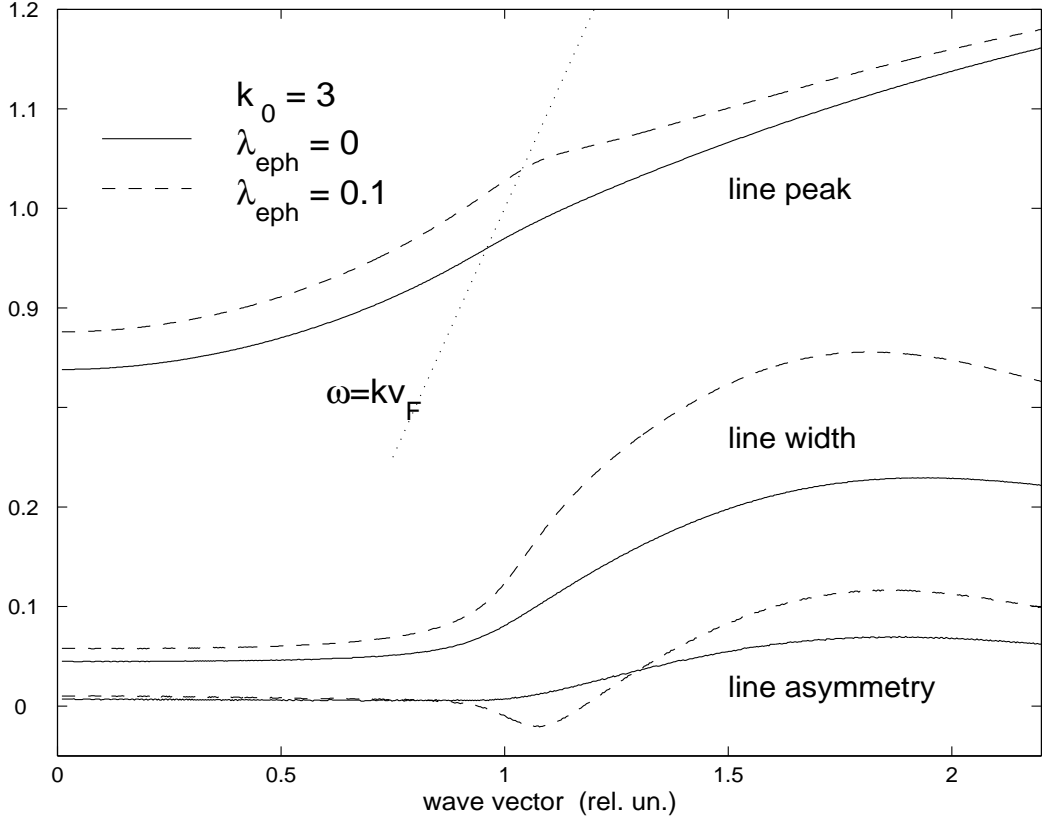}
\end{center}
\caption{Dispersion of the phonon peak (upper), 
the line width and the line asymmetry (bottom). The boundary of the
Landau damping region is shown as a dotted line.}
\end{figure}

\clearpage
\begin{figure}
\begin{center}
\epsfxsize=140mm
\epsfysize=110mm
\epsfbox{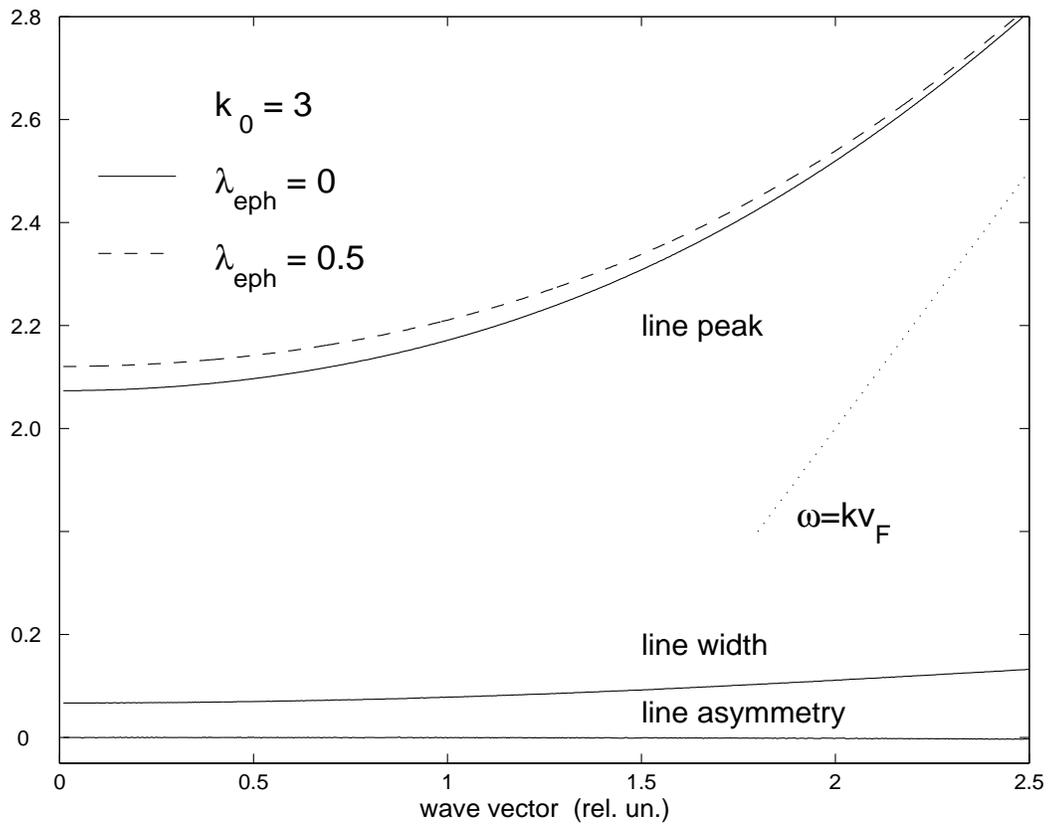}
\end{center}
\caption{Dispersion of the plasmon peak. }
\end{figure}

\end{document}